\documentclass{jnmp}

\usepackage{amsmath}

\JNMPnumberwithin{equation}{section}

\newtheorem{thm}{Theorem}[section]
\newtheorem{lem}[thm]{Lemma}

\newtheorem{pro}[thm]{Proposition}

\newtheorem{rmks}[thm]{Remarks}

\theoremstyle{definition}

\newcommand{\bbR}{\mathbb{R}}
\newcommand{\bbZ}{\mathbb{Z}}
\newcommand{\bbN}{\mathbb{N}}

\newcommand{\bbC}{\mathbb{C}}

\newcommand{\fF}{{\mathfrak{F}}}

\newcommand{\fL}{{\mathfrak{L}}}
\newcommand{\cD}{{\mathcal{D}}}
\newcommand{\cP}{{\mathcal{P}}}
\newcommand{\cK}{{\mathcal{K}}}

\newcommand{\cO}{{\mathcal{O}}}

\newcommand{\cF}{{\mathcal{F}}}

\newcommand{\Hom}{\mathrm{Hom}}

\newcommand{\cS}{{\mathcal{S}}}

\newcommand{\Vect}{\mathrm{Vect}}

\newcommand{\ord}{\mathrm{ord}}
\newcommand{\Gr}{{\mathrm{Grad}}}
\newcommand{\half}{\frac{1}{2}}

\def\a{\alpha}

\def\l{\lambda}

\begin{document}

%
\renewcommand{\evenhead}{B. Agrebaoui, N. Ben Fraj and S. Omri}
\renewcommand{\oddhead}{On the Cohomology  of the Lie Superalgebra of Contact
Vector Fields on $S^{1|2}$ }

%
\thispagestyle{empty}

\FirstPageHead{*}{*}{20**}{\pageref{firstpage}--\pageref{lastpage}}{Article}

\copyrightnote{2006}{B. Agrebaoui N. Ben Fraj and S. Omri}

\Name{On the Cohomology  of the Lie Superalgebra of Contact Vector
Fields on $S^{1|2}$ }

\label{firstpage}

\Author{B. Agrebaoui~$^\dag$ N. Ben Fraj~$^\dag$ and S.
Omri~$^\ddag$}

\Address{$^\dag$ D\'epartement de Math\'ematiques, Facult\'e des
Sciences de Sfax, Route de Soukra 3018 Sfax BP 802, Tunisie\\
~~E-mail:bagreba@fss.rnu.tn\\[10pt] $^\dag$ Institut Sup\'{e}rieur
de Sciences Appliqu\'{e}es et Technologie, Sousse, Tunisie \\
~~E-mail:~benfraj\_nizar@yahoo.fr\\[10pt] $^\ddag$ D\'epartement
de Math\'ematiques, Facult\'e des Sciences de Sfax, Route de
Soukra,\\~~3018 Sfax BP 802, Tunisie\\
~~E-mail:~omri\_salem@yahoo.fr}

\Date{Received Month *, 200*; Revised Month *, 200*; Accepted
Month *, 200*}

\begin{abstract}
We investigate the first cohomology space associated with the
embedding of the Lie superalgebra $\cK(2)$ of contact vector
fields on the (1,2)-dimensional supercircle $S^{1\mid 2}$ in the
Lie superalgebra $\cS\Psi \cD \cO(S^{1\mid 2})$ of
superpseudodifferential operators with smooth coefficients.
Following Ovsienko and Roger, we show that this space is
ten-dimensional with only even cocycles and we give explicit
expressions of the basis cocycles.
\end{abstract}

\bigskip
\thispagestyle{empty}

\section{Introduction}
V. Ovsienko and C. Roger \cite{RO1} calculated the space
$H^1(\Vect(S^1),~\Psi \cD \cO(S^1)),$ where $\Vect(S^1)$ is the
Lie algebra of smooth vector fields on the circle $S^1$ and $\Psi
\cD \cO(S^1)$ is the space of pseudodifferential operators with
smooth coefficients. The action is given by the natural embedding
of $\Vect(S^1)$ in $\Psi \cD \cO(S^1)$. They used the results of
D. B. Fuchs \cite{Fu} on the cohomology of $\Vect(S^1)$ with
coefficients in weighted densities to determine the cohomology
with coefficients in the graded module $Gr(\Psi \cD \cO(S^1))$,
namely $H^1(\Vect(S^1),~Gr^{p}(\Psi \cD \cO(S^1)))$; here
$Gr^p(\Psi \cD \cO(S^1))$ is isomorphic, as $\Vect(S^1)$-module,
to the space of weighted densities $\cF_p$ of weight $-p$ on
$S^1$. To compute $H^1(\Vect(S^1),~\Psi \cD \cO(S^1)),$ V.
Ovsienko and C. Roger applied the theory of spectral sequences to
a filtered module over a Lie algebra.

In a recent paper \cite{AB}, using the same methods as in the
paper \cite{RO1}, two of the authors computed $H^1(\cK(1),~\cS\Psi
\cD \cO(S^{1\mid 1}))$, where $\cK(1)$ is the Lie superalgebra
$\cK(1)$ of contact vector fields on the supercircle $S^{1\mid 1}$
and $\cS\Psi \cD \cO(S^{1\mid 1})$ is the space of
superpseudodifferential operators on $S^{1\mid 1}.$

Here, we follow again the same methods by V. Ovsienko and C. Roger
\cite{RO1} to calculate $H^1(\cK(2),~\cS\Psi \cD \cO(S^{1\mid
2})).$

The paper (\cite{RO1}) contains also the classification of
polynomial deformations of the natural embedding of $\Vect(S^1)$
in $\Psi \cD \cO(S^1)$. The multi-parameter deformations of the
embedding of $\cK(1)$ into $\cS\Psi\cD\cO(S^{1\mid 1})$ are
classified in (\cite{BO}). Our aim is this classification for the
case $S^{1\mid 2}$.

\section{Definitions and Notations}
Let $S^{1\mid n}$ be the supercircle with local coordinates
$(\varphi;~\theta_1,\ldots,\theta_n),$ where
$\theta=(\theta_1,\ldots,\theta_n)$ are the odd variables. More
precisely, let $x=e^{i\varphi},$ in what follows by $S^{1\mid n}$
we mean the supermanifold $(\bbC^*)^{1\mid n},$ whose underlying
is $\bbC\setminus \{0\}.$ Any contact structure on $S^{1\mid n}$
can be given by the following $1$-form:
\begin{equation*}
\a_n=dx+\sum_{i=1}^n\theta_id\theta_i.
\end{equation*}
Let $\cK(n)$ be the Lie superalgebra of vector fields on $S^{1\mid
n}$ whose Lie action on $\a_n$ amounts to a multiplication by a
function. Any element of $\cK(n)$ is of the form (see \cite{Ra})
\begin{equation*}
v_F=F\partial_x+\frac{(-1)^{p(F)+1}}{2}
\sum_{i=1}^n\eta_i(F)\eta_i,
\end{equation*}
where $F\in C^{\infty}(S^{1\mid n}),~p(F)$ is the parity of $F$
and $\eta_i=\partial_{\theta_i}-\theta_i\partial_x$. The bracket
is given by
\begin{equation*}
[v_F,\, v_G]=v_{\{F,G\}},
\end{equation*}
where
\begin{equation*}
\{F,G\}=FG'-F'G+\frac{(-1)^{p(F)+1}}{2}
\sum_{i=1}^n\eta_i(F)\eta_i(G).
\end{equation*}
The Lie superalgebra $\cK(n)$ is called the Lie superalgebra of
contact vector fields.

The superspace of the supercommutative algebra of
superpseudodifferential symbols on $S^{1\mid n}$ with its natural
multiplication is spanned by the series
\begin{equation*}
\cS\cP(n)=
\Big\{A=\sum_{k=-M}^{\infty}\sum_{\epsilon=(\epsilon_1,\ldots,\epsilon_n)}a_{k,\,
\epsilon}(x,\theta)\xi^{-k}\bar{\theta}_1^{\epsilon_1}\cdots
\bar{\theta}_n^{\epsilon_n}|\ a_{k,\, \epsilon}\in
C^{\infty}(S^{1\mid n});~\epsilon_i=0,\, 1; \, M\in\bbN \Big\},
\end{equation*}
where $\xi$ corresponds to $\partial_x$ and $\bar{\theta}_i$
corresponds to $\partial_{\theta_i}$ ($p(\bar{\theta}_i)=1)$. The
space $\cS\cP(n)$ has a structure of the Poisson Lie superalgebra
given by the following bracket:
\begin{equation*}
\{A,~B\}=\frac{\partial(A)}{\partial{\xi}}\frac{\partial(B)}{\partial{x}}-
\frac{\partial(A)}{\partial{x}}\frac{\partial(B)}{\partial{\xi}} -
(-1)^{p(A)}\sum_{i=1}^n
\Big(\frac{\partial(A)}{\partial{\theta_i}}\frac{\partial(B)}{\partial{\bar{\theta}_i}}
+\frac{\partial(A)}{\partial{\bar{\theta}_i}}\frac{\partial(B)}{\partial{{\theta}_i}}\Big).
\end{equation*}

The associative superalgebra of superpseudodifferential operators
$\cS\Psi\cD\cO(S^{1\mid n})$ on $S^{1\mid n}$ has the same
underlying vector space as $\cS\cP(n)$, but the multiplication is
now defined by the following rule:
\begin{equation*}
 A\circ B=\sum_{\alpha\geq 0,\, \nu_i=0,\, 1}\frac{(-1)^{p(A)
 +1}}{\a!}
(\partial_{\xi}^{\alpha}\partial_{\bar
\theta_i}^{\nu_i}A)(\partial_x^{\alpha}\partial_{\theta_i}^{\nu_i}B).
\end{equation*}
 This composition rule induces the supercommutator defined by:
 \begin{equation*}
 [A,~B]=A\circ B -
(-1)^{p(A)p(B)}B\circ A.
\end{equation*}
\section{The space of weighted densities on $S^{1|2}$}
Recall the definition of the  $\Vect(S^1)$-module of weighted
densities on $S^1$. Consider the $1$-parameter action of
$\Vect(S^1)$ on $C^{\infty}(S^1)$ given by
\begin{equation*}
L^{\lambda}_{X(x)\partial}(f(x))= X(x)f'(x)+\lambda X'(x)f(x),
\end{equation*}
where $f\in C^{\infty}(S^1)$ and $\l\in\bbR$. Denote $\cF_\l$ the
$\Vect(S^1)$-module structure on $C^{\infty}(S^1)$ defined by this
action. Note that the adjoint $\Vect(S^1)$-module is isomorphic to
$\cF_{-1}$. Geometrically, $\cF_\l$ is the space of weighted
densities of weight $\l$ on $S^1$, i.e., the set of all
expressions: $f(x)(dx)^{\lambda}$, where $f\in C^{\infty}(S^1)$.
We have analogous definition of weighted densities in the
supercase (see \cite{AB}) with $dx$ replaced by $\a_n.$

Consider the $1$-parameter action of $\cK(n)$ on
$C^{\infty}(S^{1|n})$ given by the rule:
\begin{equation}
\label{superaction} \fL^{\lambda}_{v_F}(G)=v_{F}(G) + \lambda
F'\cdot G,
\end{equation}
where $F,\, G\in C^{\infty}(S^{1|n}), F'\equiv \partial_{x}F.$ We
denote this $\cK(1)$-module by $\Im_{\lambda}$ and the
$\cK(2)$-module by $\fF_{\l}$. Geometrically, the space $\fF_{\l}$
is the space of all weighted densities on $S^{1|2}$ of weight
$\l$:
\begin{equation}
\label{densities} \phi=f(x,\theta)\a_{2}^{\l}, \; f(x,\theta)\in
C^{\infty}(S^{1|2}).
\end{equation}

\begin{rmks}
~1) The adjoint $\cK(2)$-module is isomorphic to $\fF_{-1}$. This
isomorphism induces a contact bracket on $C^{\infty}(S^{1|2})$
given by:
\begin{equation}
\{F,G\}=\fL^{-1}_ {
v_F}(G)=FG'-F'G+\frac{(-1)^{p(F)+1}}{2}\sum_{i=1}^2(\eta_{i}F)(\eta_{i}G).
\end{equation}
2) As a $\Vect(S^1)$-module, the space of weighted densities
$\fF_{\l}$ is isomorphic to
\begin{equation*} \cF_\l\oplus \Pi
(\cF_{\l+\half}\oplus \cF_{\l+\half})\oplus \cF_{\l+1}.
\end{equation*}
\end{rmks}
\section{The structure of $\cS\cP(2)$ as a $\cK(2)$-module }
The natural embedding of $\cK(2)$ into $\cS\cP(2)$ defined by
\begin{equation}
\label{emb}
 \pi(v_{F}) = F\xi + \frac{(-1)^{p(F)+1}}{2}
\sum_{i=1}^2\eta_i(F)\zeta_i,~~\hbox{where},~\zeta_i =
\bar\theta_i - \theta_i\xi,
\end{equation}
induces a $\cK(2)$-module structure on $\cS\cP(2).$

 Setting $\deg x=\deg \theta_i=0,~\deg \xi=\deg \bar{\theta}_i=1$
for all $i,$ we endow the Poisson superalgebra $\cS\cP(2)$ with a
$\bbZ$-grading:
\begin{equation}
\label{grading} \cS\cP(2)=\widetilde{\bigoplus}_{n\in
\bbZ}\cS\cP_{n},
\end{equation}
where $\widetilde{\bigoplus}_{n\in
\bbZ}=(\bigoplus_{n<0})\bigoplus \prod_{n\geq 0}$ ~and
\begin{equation*} \cS\cP_n=\Big\{F\xi^{-n}+G\xi^{-n-1}\bar\theta_1 +
H\xi^{-n-1}\bar\theta_2 + T\xi^{-n-2}\bar\theta_1\bar\theta_2~|~
F,\, G,~H,~T \in C^{\infty}(S^{1|2})\Big\}
\end{equation*}
\noindent is the homogeneous subspace of degree
$-n$.
\bigskip
Each element of $ \cS\Psi \cD \cO(S^{1|2})$  can be expressed as
\begin{equation*}
A=\sum_{k\in \bbZ}(F_k+G_k\xi^{-1}\bar\theta_1
+H_k\xi^{-1}\bar\theta_2 +T_k\xi^{-2}\bar\theta_1\bar\theta_2
)\xi^{-n},
\end{equation*} where $F_k,\, G_k,~H_k,~T_k\in
C^{\infty}(S^{1|2})$. We define the {\it order} of $A$ to be
\begin{equation*} \ord(A)=\sup\{k~|~ F_k\neq 0 \hbox{ or } G_k\neq 0
\hbox{ or } H_k\neq 0 \hbox{ or }T_k\neq 0\}.
\end{equation*}
This definition of order equips $ \cS\Psi \cD \cO(S^{1|2})$ with a
decreasing filtration as follows: \noindent set
\begin{equation*}
{\bf F}_n=\{A\in \cS\Psi \cD \cO(S^{1|2}),\,\ord(A)\leq
-n\},
\end{equation*}
where $n\in \bbZ$. So one has
\begin{equation}
\label{filtration} \ldots\subset {\bf F}_{n+1}\subset {\bf
F}_n\subset \ldots
\end{equation}
This filtration is compatible with the multiplication and the
Poisson bracket, that is, for $A\in {\bf F}_n$ and $B\in {\bf
F}_m$, one has $A\circ B\in {\bf F}_{n+m}$ and $\{A,B\}\in {\bf
F}_{n+m-1}$. This filtration makes $\cS\Psi \cD \cO(S^{1|2})$ an
associative filtered superalgebra. Consider the associated graded
space
\begin{equation*}
Gr(\cS\Psi \cD \cO(S^{1|2}))=\widetilde{\bigoplus}_{n\in \bbZ}{\bf
F}_n/{\bf F}_{n+1}.
\end{equation*}
\noindent The filtration (\ref{filtration}) is also compatible
with the natural action of $\cK(2)$ on $\cS\Psi \cD \cO(S^{1|2}).$
Indeed, if $v_F\in \cK(2)$ and $A\in {\bf F}_n$, then
\begin{equation*}
v_F(A)=[v_F,A]\in {\bf F}_{n}.
\end{equation*}
The induced $\cK(2)$-module on the quotient ${\bf F}_n/{\bf
F}_{n+1}$ is isomorphic to the $\cK(2)$-module $\cS\cP_n$.
 Therefore, the $\cK(2)$-module $Gr(\cS\Psi \cD \cO(S^{1|2}))$, is isomorphic
to the graded $\cK(2)$-module $\cS\cP(2)$, that is
\begin{equation*}
\cS\cP(2)\simeq \widetilde{\bigoplus}_{n\in \bbZ}{\bf F}_n/{\bf
F}_{n+1}.
\end{equation*}

Recall that a $C^{\infty}$ function on $S^{1|2}$ has the form
$F=f_0 + f_1\theta + f_2\theta + f_{12}\theta_1\theta_2$ with
$f_0, f_1, f_2, f_{12}\in C^{\infty}(S^1)$ and a $C^{\infty}$
function on $S_i^{1|1} (i=1, 2),$ where $S_i^{1\mid 1}$ is the
supercircle with local coordinates $(\varphi, \theta_i),$ has the
form $F=f_0 + f_i\theta_i~(f_{12}=f_{3-i}=0)$ with $f_0, f_i\in
C^{\infty}(S^1).$ Then the Lie superalgebra $\cK(2)$ has two
subsuperalgebras $\cK(1)_i \hbox{ for }~i=1,~2$ isomorphic to
$\cK(1)$ defined by
\begin{equation*}
 \cK(1)_i =\Big\{v_F=F\partial_x+\frac{(-1)^{p(F)+1}}{2}
\sum_{i=1}^2\eta_i(F)\eta_i~|~F\in C^{\infty}(S_i^{1|1})\Big\}.
\end{equation*}
Therefore, $\cS\cP(2)$ and $\fF_{\lambda}$ are
$\cK(1)_i$-modules.\\ For $i=1, 2,$ let $\Im_{\lambda}^i$ be the
$\cK(1)_i$-module of weighted densities of weight $\lambda$
on~$S_i^{1\mid 1}.$
\begin{pro}
\label{prop1}~~ 1) As a $\cK(1)_i$-module, $i=1,~2,$ we have
\begin{equation*}
\renewcommand{\arraystretch}{1.4}
\cS\cP_n\simeq {\fF}_n\oplus \Pi({\fF}_{n+\half}\oplus
{\fF}_{n+\half})\oplus {\fF}_{n+1} \hbox{ for }~n=0,-1.
\end{equation*}
\medskip
2) For $n\neq 0,-1:$\\
\medskip
a) The following subspace of $\cS\cP_{n}:$
\begin{equation}
\label{deci} \cS\cP_{n,~i} = \left\{
\begin{array}{lll}
 B_F^{(n,i)}&=& F \theta_{3-i}\bar\theta_{3-i}\xi^{-n-1} + \theta_{3-i}(\eta_{3-i}-
  \frac{1}{2}\eta_i)(F)\zeta_i \zeta_{3-i}\xi^{-n-2} \mid \\[6pt]&& F\in C^{\infty}(S^{1\mid 2})
\end{array}
 \right\}
\end{equation}
is a $\cK(1)_i$- module, $i=1, 2,$ isomorphic to $\fF_{n+1}.$\\
\medskip
b) As a $\cK(1)_i$-module we have
\begin{equation*}
\renewcommand{\arraystretch}{1.4}
\cS\cP_n/\cS\cP_{n,~i}\simeq {\fF}_n\oplus \Pi
({\fF}_{n+\half}\oplus {\fF}_{n+\half}),~~i=1, 2.
\end{equation*}
\end{pro}

\begin{proofname}.
First, note that for $n=0,~-1,$ the $\cK(1)_i$-module $\cS\cP_n$
with the grading (\ref{grading}) is the direct sum of four
$\cK(1)_i$-modules, $i=1, 2.$

For $n=0,$ the four $\cK(1)_i$-modules are
\begin{equation*}
\label{dec0}
\renewcommand{\arraystretch}{1.4}
\begin{array}{lll}
\cS\cP_{(0,~0)} &=& \left\{A_F^{(0,~0)}= F\mid~ F\in
C^{\infty}(S^{1\mid 2})\right\}\ ,\\[10pt]
\medskip
\cS\cP_{(0,~\half,~i)}&=&\left\{
\begin{array}{lll}
A_F^{(0,~\half,~i)}&=& \theta_i F - (1 -
2\theta_{3-i}\partial_{\theta_{3-i}})(F)\bar{\theta}_i \xi^{-1}~
-\\&& \theta_{3-i}\partial_{\theta_i}( F)\bar{\theta}_{3-i}
\xi^{-1} + F'\theta_{3-i} \bar{\theta}_i \bar{\theta}_{3-i}
\xi^{-2}\mid\\ &&F\in C^{\infty}(S^{1\mid 2})
\end{array}
\right\}, \\[10pt]
\medskip
\widetilde{\cS\cP}_{(0,~\half,~i)}^{} &=& \left\{
\begin{array}{lll}
\widetilde{A}_F^{(0,~\half,~i)}&=&
\theta_i(\partial_{\theta_{3-i}} - 2\partial_{\theta_i} +
2\theta_{3-i}\partial_{\theta_{3-i}}\partial_{\theta_i})(F)
\bar\theta_{3-i} \xi^{-1}~ +
\\ && \frac{1}{2}(3F - (-1)^{p(F)}F)
\bar\theta_{3-i}\xi^{-1}~ + \\ && (-1)^{p(F)}
(\partial_{\theta_{3-i}}-
\partial_{\theta_i}+ \theta_i\partial_x)(F)\bar\theta_i \bar\theta_{3-i}\xi^{-2} \mid\\&&
F\in C^{\infty}(S^{1\mid 2})
\end{array}
\right\},\\[10pt]
\medskip
\cS\cP_{(0,~1,~i)} &=& \left\{
\begin{array}{lll}
A_F^{(0,~1,~i)}&=& F \theta_{3-i}\bar\theta_{3-i}\xi^{-1}  +
\theta_{3-i} (\eta_{3-i} - \frac{1}{2} \eta_i)(F)\zeta_i
\zeta_{3-i}\xi^{-2} \mid \\&& F\in C^{\infty}(S^{1\mid 2})
\end{array}
\right\}.
\end{array}
\end{equation*}
\par For $n = -1,$ the four $\cK(1)_i$-modules are
\begin{equation*}
\label{dec1}
\begin{array}{lll}
\cS\cP_{(-1,~0)} &=& \left\{A_F^{(-1,~0)} = F \xi +
\frac{(-1)^{p(F) + 1}}{2}\Big(\eta_1 (F)\zeta_1 +\eta_2
(F)\zeta_2\Big)  \mid ~F\in C^{\infty}(S^{1\mid 2})\right\},
\\[10pt]
\medskip
\cS\cP_{(-1,~ \half,~i)} &=&\left\{
\begin{array}{lll}
A_F^{(-1,~\half,~i)} &=& F\zeta_i -(\theta_{3-i}\eta_i +
\theta_i\partial_{\theta_{3-i}})(F)\bar\theta_{3-i}~- \\[6pt] &&
(-1)^{p(F)}\partial_{\theta_{3-i}}(F)
\bar\theta_i\bar\theta_{3-i}\xi^{-1} \mid~F\in C^{\infty}(S^{1\mid
2})
\end{array}
\medskip
\right\},\\[10pt]
\widetilde{\cS\cP}_{(-1,~\half,~i)}&=&\left\{\widetilde{A}_F^{(-1,~\half,~i)}
= F\zeta_i + (1 - \theta_{3-i}\eta_i)(F)\bar\theta_{3-i}\mid ~F\in
C^{\infty}(S^{1\mid 2})\right\},\\[10pt]
\medskip
\cS\cP_{(-1,~1,~i)}&=&\left\{
\begin{array}{lll}
A_F^{(-1,~1,~i)}&=& F\theta_{3-i} \bar\theta_{3-i} + \theta_{3-i}
(\eta_{3-i}- \frac{1}{2}\eta_i)(F)\zeta_i\zeta_{3-i}\xi^{-1} \mid
\\[6pt]&& F\in C^{\infty}(S^{1\mid 2})
\end{array}
\right\}.
\end{array}
\end{equation*}
The action of $\cK(1)_i$ on $\cS\cP_{(n,~0)}$ and on
$\cS\cP_{(n,~1,~i)}$ for $n=0,~-1$ is induced by the embedding
(\ref{emb}) as follows
\begin{equation*}
\begin{array}{lll}
 v_G\cdot A_F^{(n,~0)}&=\Big\{\pi(v_G),~A_F^{(n,~0)}\Big\}\\
 &=A_{\fL^{n}_{v_G}(F)}^{(n,~0)}
  \end{array}
  \quad
\text{ and }\quad
\begin{array}{lll}
 v_G\cdot A_F^{(n,~1,~i)}&=\Big\{\pi(v_G),~A_F^{(n,~1,~i)}\Big\}\\
 &=A_{\fL^{n+1}_{v_G}(F)}^{(n,~1,~i)},
\end{array}
\end{equation*}
where $F\in C^{\infty}(S^{1\mid 2})$ and $G\in
C^{\infty}(S_{i}^{1\mid 1}).$ Therefore, the natural maps
\begin{equation}
\label{inj1}
\begin{array}{lcll} \psi_{n,~0}^i :&\fF_{n} &\longrightarrow
&\cS\cP_{(n,~0)}\\[6pt] &F\a_2^{n}&\longmapsto &A_F^{(n,~0)}
\end{array}
 \quad
\text{ and }\quad
\begin{array}{lcll}
 \psi_{n,~1}^i :&\fF_{n+1} &\longrightarrow
 &\cS\cP_{(n,~1,~i)}\\[6pt] &F\a_2^{n+1}&\longmapsto&A_F^{(n,~1,~i)}
\end{array}
\end{equation}
provide us with isomorphisms of $\cK(1)_i$-modules, $i=1,~2.$

\noindent The action of $\cK(1)_i$ on $\cS\cP_{(n,~\half,~i)}$ and
on $\widetilde{\cS\cP}_{(n,~\half,~i)}$ for $n=0,~-1$ is given by
\begin{equation*}
\begin{array}{lll}
 v_G\cdot A_F^{(n,~\half,~i)}&=\Big\{\pi(v_G),~A_F^{(n,~\half,~i)}\Big\}\\
 &=A_{\fL^{n+\half}_{v_G}(F)}^{(n,~\half,~i)}
  \end{array}
  \quad
\text{ and }\quad
\begin{array}{lll}
 v_G\cdot \widetilde{A}_F^{(n,~\half,~i)}&=\Big\{\pi(v_G),~\widetilde{A}_F^{(n,~\half,~i)}\Big\}\\
 &=\widetilde{A}_{\fL^{n+1}_{v_G}(F)}^{(n,~\half,~i)},
\end{array}
\end{equation*}
where $F\in C^{\infty}(S^{1\mid 2})$ and $G\in
C^{\infty}(S_{i}^{1\mid 1}).$ Therefore, the natural maps
\begin{equation}
\label{inj2}
\begin{array}{lll} \psi_{n,~\half}^i :\Pi(\fF_{n+\half}) &\longrightarrow
&\cS\cP_{(n,~\half,~i)}\\[6pt]
\quad\quad\Pi(F\a_2^{n+\half})&\longmapsto &A_F^{(n,~\half,~i)}
\end{array}
 \quad
\text{ and }\quad
\begin{array}{lcll}
 \widetilde{\psi}_{n,~\half}^i :\Pi(\fF_{n+\half}) &\longrightarrow
 &\widetilde{\cS\cP}_{(n,~\half,~i)}\\[6pt] \quad\quad\Pi(F\a_2^{n+\half})
 &\longmapsto&\widetilde{A}_F^{(n,~\half,~i)}
\end{array}
\end{equation}
provide us with isomorphisms of $\cK(1)_i$-modules.

Second, for $n\neq 0,~-1,$ the action of $\cK(1)_i$ on
$\cS\cP_{n,~i}$ is given by
\begin{equation*}
 v_G\cdot B_F^{(n,~i)}=\Big\{\pi(v_G),~B_F^{(n,~i)}\Big\} = B_{\fL^{n+1}_{v_G}(F)}^{(n,~i)},
\end{equation*}
where $F\in C^{\infty}(S^{1\mid 2})$ and $G\in
C^{\infty}(S_{i}^{1\mid 1}).$ Therefore,
$\cS\cP_{n,~i}\simeq\fF_{n+1}$ as a $\cK(1)_i$-module. The induced
$\cK(1)_i$-module on the quotient $\cS\cP_n/\cS\cP_{n,~i}$ has the
direct sum decomposition of the three $\cK(1)_i$- modules,~
$\cS\cP_{(n,~0,~i)},~\cS\cP_{(n,~\half,~i)}$ ~and~
$\widetilde{\cS\cP}_{(n,~\half,~i)},$ defined by
\begin{equation*}
\label{decn}
\begin{array}{lll}
\cS\cP_{(n,~0~i)} &=& \left\{
\begin{array}{lll}
A_F^{(n,~0~i)} &=& F \xi^{-n} + \frac{(-1)^{p(F)}}{2}(
\frac{1}{2n+1}\theta_{3-i}\eta_{3-i}\eta_i~-
\eta_i)(F)\zeta_i\xi^{-n-1}~+
\\[10pt]&&
(\partial_{\theta_{3-i}} +
\frac{3n+1}{2n+1}\theta_i\partial_{\theta_{3-i}}\partial_{\theta_i})(F)
\bar\theta_{3-i}\xi^{-n-1}~+ \\[10pt] &&
 \frac{n+1}{2n+1}(\theta_{3-i}\eta_i^3 +
\eta_{i}\eta_{3-i})(F)\bar\theta_{3-i}\bar\theta_i\xi^{-n-2}\mid
\\[10pt]&& F\in C^{\infty}(S^{1\mid 2})
\end{array}
\right\},\\[1.5cm] \cS\cP_{(n,~\half,~i)}^{} &=&\left\{
\begin{array}{lll}
A_F^{(n,~\half,~i)} &=&(\theta_{3-i}\partial_{\theta_{3-i}} -1)
(F)\zeta_i\xi^{-n-1}~+
\\[10pt]
&& \frac{1}{2n+1}(n\theta_i\theta_{3-i}\partial_x -
\theta_{3-i}\partial_{\theta_i})(F)\bar\theta_{3-i}
\xi^{-n-1}~+\\[10pt] &&
\frac{n+1}{2n+1}F'\theta_{3-i}\bar\theta_i\bar\theta_{3-i}\xi^{-n-2}\mid~~F\in
C^{\infty}(S^{1\mid 2})
\end{array}
\right\},\\[1.5cm] \widetilde{\cS\cP}_{(n,~\half,~i)} &=& \left\{
\begin{array}{lll}
\widetilde{A}_F^{(n,~\half,~i)}&=& (-1)^{p(F)}\theta_{3-i}(1 +
\theta_i\partial_{\theta_{3-i}} -
\frac{n}{2n+1}\theta_i\partial_{\theta_i})(F)\xi^{-n}~+
\\[10pt] && (\theta_{3-i}\partial_{\theta_{3-i}} -
\frac{n}{2n+1}\theta_{3-i}\eta_i)(F)\bar\theta_i\xi^{-n-1}~+\\[10pt]
&& (-1)^{p(F)}(\theta_{3-i}\partial_x + \eta_{3-i})(F)\zeta_i
\bar\theta_{3-i}\xi^{-n-2}\mid\\[10pt]&& F\in C^{\infty}(S^{1\mid
2})
\end{array}
\right\}.
\end{array}
\end{equation*}
The action of $\cK(1)_i$ on $\cS\cP_{(n,~j,~i)}$ and on
$\widetilde{\cS\cP}_{(n,~\half,~i)}$ is induced by the the action
of $\cK(1)_i$ on $\cS\cP_n/\cS\cP_{n,~i}$ and a direct computation
shows that one has:
\begin{equation*}
 v_G\cdot A_F^{(n,~j,~i)}= A_{\fL^{n+j}_{v_G}(F)}^{(n,~j,~i)}
\quad \text{for}\quad j=0,~\half \quad \text{and }\quad
 v_G\cdot \widetilde{A}_F^{(n,~\half,~i)}=
 \widetilde{A}_{\fL^{n+\half}_{v_G}(F)}^{(n,~\half,~i)},
\end{equation*}
where $F\in C^{\infty}(S^{1\mid 2})$ and $G\in
C^{\infty}(S_{i}^{1\mid 1}), i=1,2.$ Therefore, the natural maps
\begin{equation*}
\begin{array}{lll} \psi_{n,~0}^i :&\fF_n \longrightarrow
&\cS\cP_{(n,~0,~i)}\\[6pt] &F\a_2^n \longmapsto &A_F^{(n,~0,~i)}
\end{array}
 \quad
\text{ , }\quad
\begin{array}{lll}
 \psi_{n,~\half}^i :\Pi(\fF_{n+\half}) &\longrightarrow
 &\cS\cP_{(n,~\half,~i)}\\[6pt] \quad\quad\Pi(F\a_2^{n+\half})
 &\longmapsto &A_F^{(n,~\half,~i)}
\end{array}
\end{equation*}
\begin{equation}
\label{inj3} \quad \text{ and }\quad
\begin{array}{lll}
 \widetilde{\psi}_{n,~\half}^i :\Pi(\fF_{n+\half}) &\longrightarrow
 &\widetilde{\cS\cP}_{(n,~\half,~i)}\\[6pt] \quad\quad\Pi(F\a_2^{n+\half})
 &\longmapsto&\widetilde{A}_F^{(n,~\half,~i)}
\end{array}
\end{equation}
provide us with isomorphisms of $\cK(1)_i$-modules. This completes
the proof.
\end{proofname}
\section{The first cohomology space $H^1(\cK(2),\cS\cP(2))$}
Let us first recall some fundamental concepts from cohomology
theory~(\cite{Fu}). Let $\mathfrak{g}=\mathfrak{g}_0\oplus \mathfrak{g}_1$ be
a Lie superalgebra acting on a super vector space $V=V_0\oplus
V_1$. The space $\Hom(\mathfrak{g},\, V)$ is $\bbZ_2$-graded via
\begin{equation}
\label{grade} \Hom(\mathfrak{g}, V)_b=\displaystyle \oplus_{a\in
\bbZ_2}\Hom(\mathfrak{g}_a, V_{a+b}); \; b\in \bbZ_2.
\end{equation}
 According to the $\bbZ_2$-grading (\ref{grade}), each $c\in  Z^1(\mathfrak{g}, V)$, is
 broken to
 $(c',c'')\in \Hom(\mathfrak{g}_0,\, V)\oplus \Hom(\mathfrak{g}_1,\, V)$
 subject to the following three equations:
\begin{equation}
\label{cocycle}
\begin{array}{lllll}
 (E_1)\qquad c'([g_1,g_2]) - g_1.c'(g_2) + g_2.c'(g_1)&=& 0
 &\hbox{ for any }  &g_1,g_2\in\mathfrak{g}_{0}, \\[10pt]
 (E_2)\qquad c''([g,h]) - g.c''(h) + h.c'(g)&=& 0 &\hbox{ for any }
 &g\in\mathfrak{g}_0,h\in\mathfrak{g}_{1},\\[10pt]
 (E_3)\qquad
 c'([h_1,h_2]) - h_1c''(h_2) - h_2c''(h_1)&=&0 &\hbox{ for any }
 &h_1,h_2\in\mathfrak{g}_{1}.
 \end{array}
\end{equation}
\begin{pro}
\label{cohomdensities}

1)
\begin{equation*}
H^1(\cK(1)_i,\fF_{\lambda})_0\simeq\left\{
\begin{array}{ll}
\bbR^3&\makebox{ if }~\l=0,\\ \bbR &\makebox{ if }~\l=1, \\
0&\makebox { otherwise }.
\end{array}
\right.\end{equation*}
The respective nontrivial $1$-cocycles are
\begin{equation}
\label{C0123}
\begin{array}{ll}
 \displaystyle C_0(v_F) =\displaystyle
\frac{1}{4}(3F+(-1)^{p(F)}F),~\displaystyle C_1(v_F)
=\displaystyle F',~\displaystyle
C_2(v_F)=\bar{\eta}_i(F')\theta_{3-i} &\hbox{ if } \l=0,\\
\\[5pt]
\displaystyle C_3(v_F)=\bar{\eta}_i(F'')\theta_{3-i} &\hbox{ if }
\l=1,
\end{array}
\end{equation}
where $\bar{\eta}_i = \partial_{\theta_i} + \theta_i\partial_x$,
$v_F\in\cK(1)_i$ and $F=f_0+f_i\theta_i$.\\
\medskip
2)
\begin{equation*}
H^1(\cK(1)_i,\fF_{\lambda})_1\simeq\left\{
\begin{array}{lll} \bbR
&\makebox{ if }~\l=\frac{1}{2},~\frac{3}{2}, \\ \bbR^2& \makebox{
if }~\l=-\frac{1}{2}, \\ 0& \makebox { otherwise}.
\end{array}
\right.
\end{equation*}
 It is spanned by the following
$1$-cocycles:
\begin{equation}
\label{C4567} \left\{
\begin{array}{ll}
 \displaystyle C_4(v_F) = \displaystyle
\frac{1}{4}(3F+(-1)^{p(F)}F)\theta_{3-i}, ~~~\displaystyle
C_5(v_F) = \displaystyle F'\theta_{3-i} &\hbox{ if }
\l=-\frac{1}{2},\\[10pt] C_6(v_F) = \bar{\eta}_i(F') &\hbox{ if }
\l=\frac{1}{2},
\\[10pt]
\displaystyle C_7(v_F) = \bar{\eta}_i(F'') &\hbox{ if }
\l=\frac{3}{2}.
\end{array}
\right.
\end{equation}
\end{pro}

To prove Proposition \ref{cohomdensities}, we need the following
result (see \cite{AB}).
\begin{pro}{\cite{AB}}
\label{prop2}

1) The  space $H^1(\cK(1)_i,\Im_{\lambda}^i)_0, i=1, 2,$ has the
following structure:
\begin{equation*}
H^1(\cK(1)_i,\Im_{\lambda}^i)_0\simeq\left\{\begin{array}{ll}
\bbR^2 &\makebox{ if }\l=0, \\ 0 &\makebox { otherwise}.
\end{array}
\right.
\end{equation*}

The space $H^1(\cK(1)_i,\Im^i_0)_0$ is generated  by the
cohomology classes of the $1$-cocycles
\begin{equation}
\label{Cc}
 c_0(v_F)=\frac{1}{4}(3F+(-1)^{p(F)}F)~
\makebox{ and }~c_1(v_F)=F'.
\end{equation}
2)\begin{equation*}
H^1(\cK(1)_i,\Im^i_{\lambda})_1\simeq\left\{\begin{array}{ll} \bbR
&\makebox{ if }\l=\half, \frac{3}{2},\\ 0 &\makebox { otherwise}.
\end{array}
\right.
\end{equation*}
It is spanned  by the nontrivial $1$-cocycles
\begin{equation}
\left\{
\begin{array}{ll}
 c_2(v_F)=\bar\eta_i(F') &\makebox{ if } \l=\half,\\[10pt]
 c_3(v_F)=\bar\eta_i(F'') &\makebox{ if }
\l=\frac{3}{2}.
\end{array}
\right.
\end{equation}

\end{pro}
\noindent{\it Proof of Proposition \ref{cohomdensities}}: Let
$F\a_2^{\lambda}=(f_0+f_1\theta_1+
f_2\theta_2+f_{12}\theta_1\theta_2)\a_2^{\lambda}
\in\fF_{\lambda}$. The map
\begin{equation*}
\begin{array}{lcll} \Phi:&\fF_{\lambda} &\longrightarrow&
\Im_{\lambda}^i\oplus\Im_{\lambda+\half}^i\\
&F\a_2^{\lambda}&\longmapsto&((1-\theta_{3-i}\partial_{{\theta}_{3-i}})(F)
\a_{1,i}^{\lambda},~
(-1)^{p(F)+1}\partial_{\theta_{3-i}}(F)\a_{1,i}^{\lambda+\half}),
\end{array}
\end{equation*}
where $\a_{1,i}=dx+\theta_i d\theta_i, i=1, 2, $ provides us with
an isomorphism of $\cK(1)_i$-modules. This map induces the
following isomorphism between cohomology spaces:
\begin{equation*}
H^1(\cK(1)_i,~\fF_{\lambda})\simeq H^1(\cK(1)_i,~\Im^i_{\lambda})
\oplus H^1(\cK(1)_i,~\Im^i_{\lambda+\half}).
\end{equation*}
We deduce from this isomorphism and Proposition \ref{prop2}, the
$1$-cocycles (\ref{C0123}--\ref{C4567}). \hfill $\Box$

The space $H^1(\cK(2),~\cS\cP(2))$ inherits the grading
(\ref{grading}) of $\cS\cP(2)$, so it suffices to compute it in
each degree. The main result of this section is the following.
\begin{thm}
\label{mainth1} The space $H^1(\cK(2),~\cS\cP_n)$ is purely even.
It has the following structure:
\begin{equation*}
H^1 (\cK(2) , \cS\cP_n) \simeq \left\{
\begin{array}{ll}
\bbR^3 &\makebox{ if }~ n= -1\\ \bbR^6 &\makebox{ if }~ n= 0\\
\bbR &\makebox{ if }~ n= 1\\ 0 &\makebox{ otherwise }.
\end{array}
\right.
\end{equation*}
For $ n= -1,$ the nontrivial $1$-cocycles  are:
\renewcommand{\arraystretch}{1.4}
$$
\begin{array}{lll}
\Upsilon_1(v_F) &=& \eta_1\eta_2(F)\xi^{-1}\zeta_1\zeta_2,\\
\Upsilon_2(v_F) &=& F'\xi^{-1}\zeta_1\zeta_2,\\
\Upsilon_3(v_F)&=&\Big(\displaystyle\frac{1}{4}(F+(-1)^{p(F)+1}F)+
 \eta_2\eta_1(F\theta_1\theta_2)\Big)\xi^{-1}\zeta_1\zeta_2,
\end{array}
$$
For $ n= 0,$ the nontrivial $1$-cocycles  are:
\renewcommand{\arraystretch}{1.4}
$$
\begin{array}{lll}
\Upsilon_4(v_F) &=&\displaystyle
\frac{1}{4}(F+(-1)^{p(F)+1}F)+\eta_2\eta_1(F\theta_1\theta_2),\\
\Upsilon_5(v_F) &=& F',\\ \Upsilon_6(v_F)&=&\eta_1\eta_2(F),\\
\Upsilon_{7}(v_F)&=& (-1)^{p(F)}\Big(\eta_1(F')\zeta_1 +
\eta_2(F')\zeta_2\Big)\xi^{-1},\\
\Upsilon_{8}(v_F)&=&F''\xi^{-2}\zeta_1\zeta_2 +(-1)^{p(F)}\Big(
\eta_2(F')\zeta_1 -\eta_1(F')\zeta_2\Big)\xi^{-1},\\
\Upsilon_{9}(v_F)&=&\eta_1\eta_2(F')\xi^{-2}\zeta_1\zeta_2,
\end{array}
$$
For $ n= 1,$ the nontrivial $1$-cocycle  is:
\renewcommand{\arraystretch}{1.4}
$$
\Upsilon_{10}(v_F)=\displaystyle\frac{2}{3}F'''\xi^{-3}\zeta_1\zeta_2
+(-1)^{p(F)}\Big( \eta_2(F'')\zeta_1 -
\eta_1(F'')\zeta_2\Big)\xi^{-2}+ \displaystyle
2\eta_1\eta_2(F')\xi^{-1}. $$
\end{thm}
To prove Theorem~\ref{mainth1}, we need first to proof the
following lemma:
\begin{lem}
\label{lemma1}
 Let $C$
 be a even (resp. odd) $1$-cocycle from $\cK(2)$ to $\cS\cP_n,~n\in \bbZ$.
   If its restriction
  to $\cK(1)_1$ and to $\cK(1)_2$ is a coboundary,
 then $C$ is a coboundary.
\end{lem}
\noindent {\bf Proof.} Let $C$ be a even (resp. odd) $1$-cocycle
of $\cK(2)$ with coefficients in $\cS\cP_{n}$ such that its
restriction to $\cK(1)_1$ and to $\cK(1)_2$ is a coboundary. Using
the condition of a  $1$-cocycle, we prove that there exists $G\in
\cS\cP_{n}$ such that
\begin{equation*}
C(v_{f_0+f_i\theta_i})=\{v_{f_0+f_i\theta_i}~,~G\} \hbox{ for any
} f_0,~f_i\in C^{\infty}(S^1) \hbox{ and } i=1,~2 \end{equation*}
and
\begin{equation*}
C(v_{f_{12}\theta_1\theta_2})=\{v_{f_{12}\theta_1\theta_2}~,~G\}
\hbox{ for any } f_{12}\in C^{\infty}(S^1).
\end{equation*}
\noindent We deduce that $C(v_F)=\{v_F~,~G\},~\hbox{for any}~ F\in
C^{\infty}(S^{1\mid 2}),$ and therefore $C$ is a coboundary of
$\cK(2)$. \hfill $\Box$
\bigskip

\noindent{\it Proof of Theorem~\ref{mainth1} }: According to
Lemma~\ref{lemma1}, the restriction of any nontrivial $1$-cocycle
of $\cK(2)$ with coefficients in $\cS\cP_{n}$ to $\cK(1)_1$ or to
$\cK(1)_2$ is a nontrivial $1$-cocycle.

Using Proposition~\ref{prop1} and
Proposition~\ref{cohomdensities}, we obtain:
\begin{equation*}
H^1 (\cK(1)_i , \cS\cP_n)\simeq \left\{
\begin{array}{lll}
\bbR^7 &\hbox{ if } n=-1\\ \bbR^6 &\hbox{ if } n=0.
\end{array}
\right.
\end{equation*}
In the case $n=-1,$ the space $H^1 (\cK(1)_i, \cS\cP_{-1})$ is
spanned by the following $1$-cocyles:
\begin{gather*}
\beta^i_{l}(v_F)=\psi^i_{-1,~1}(C_l(v_F)),~~l=0,~1,~2,\\
\beta^i_{4}(v_F)=\psi^i_{-1,~\half}(\Pi(C_4(v_F))),\\
\widetilde{\beta}^i_{4}(v_F)=\widetilde{\psi}^i_{-1,~\half}(\Pi(C_4(v_F))),\\
\beta^i_{5}(v_F)=\psi^i_{-1,~\half}(\Pi(C_5(v_F))),\\
\widetilde{\beta}^i_{5}(v_F)=\widetilde{\psi}^i_{-1,~\half}(\Pi(C_5(v_F))).
\end{gather*}
In the case $n=0,$ the space $H^1 (\cK(1)_i , \cS\cP_{0})$ is
spanned by the following $1$-cocyle:
\begin{gather*}
\beta^i_{l+6}(v_F) = \psi^i_{0,~0}(C_l(v_F)),~~l=0,~1,~2,\\
\beta^i_{9}(v_F)~~ = \psi^i_{0,~1}(C_3(v_F)),\\ \beta^i_{10}(v_F)~
= \psi^i_{0,~\half}(\Pi(C_6(v_F))),\\
\widetilde{\beta}^i_{10}(v_F)~ =
\widetilde{\psi}^i_{0,~\half}(\Pi(C_6(v_F))),
\end{gather*}
where the cocycles $C_{0},\cdots,C_{6}$ are defined by the
formulae (\ref{C0123})--(\ref{C4567}) and
$\psi_{n,~j}^i,~\widetilde{\psi}_{n,~j}^i$ are as in
(\ref{inj1})--(\ref{inj2}).

According to the same propositions, we obtain $H^1 (\cK(1)_i ,
\cS\cP_n /\cS\cP_{n,~i})$ and $H^1 (\cK(1)_i , \cS\cP_{n,~i})$ for
$n\neq 0, -1$ and $i=1,~2$. By direct computations, one can now
deduce $H^1 (\cK(1)_i , \cS\cP_n)$.

Second, note that any nontrivial $1$-cocycle of $\cK(2)$ with
coefficients in $\cS\cP_{n}$ should retain the following general
form: $ \Upsilon= \Upsilon^0 + \Upsilon^1 + \Upsilon^2 +
\Upsilon^3$ where
$\Upsilon^0:\Vect(S^1)\longrightarrow\cS\cP_n,~\Upsilon^1,\Upsilon^2:
\cF_{-\half}\longrightarrow\cS\cP_n$~and $\Upsilon^3:
\cF_0\longrightarrow\cS\cP_n$ are linear maps. The space $H^1
(\cK(1)_i , \cS\cP_n) , i=1,~2,$ determines the linear maps
$\Upsilon^0,~\Upsilon^1$ and $\Upsilon^2.$  The $1$-cocycle
conditions determines $\Upsilon^3.$ More precisely, we get:

For $n=-1,$ the space $H^1 (\cK(2) , \cS\cP_{-1})$ is generated by
the nontrivial cocycles $\Upsilon_1,~\Upsilon_2$ and $~\Upsilon_3$
corresponding to the cocycles $\beta^i_2,~\beta^i_5$ and $
~\beta^i_4,$ respectively, via their restrictions to $\cK(1)_i.$

For $n=0,$ the space $H^1 (\cK(2) , \cS\cP_{0})$ is spanned by the
nontrivial cocycles
$\Upsilon_4,\Upsilon_5,\Upsilon_6,\widetilde{\Upsilon}_7,$
$\widetilde{\Upsilon}_8 \hbox{ and } \Upsilon_9$  corresponding to
the cocycles
$\beta^i_6,~\beta^i_7,~\beta^i_8,~\beta_{10}^i,~\widetilde{\beta}_{10}^i
\hbox{ and } \beta_9^i,$ respectively, via their restrictions to
$\cK(1)_i,$ where $\widetilde{\Upsilon}_{7}=\Upsilon_7 +
\Upsilon_{9} \hbox{ and } \widetilde{\Upsilon}_{8}=\Upsilon_8 +
\Upsilon_{6}.$

Finally, for $n=1,$ the space $H^1 (\cK(2) , \cS\cP_{1})$ is
generated by the nontrivial cocycle $\Upsilon_{10}$  corresponding
to the cocycle $\psi_{1,~0}^i(C_3(v_F))$ with $\psi_{1,~0}^i$ as
in (\ref{inj3}) via its restriction to $\cK(1)_i.$

Theorem \ref{mainth1} is proved.\hfill $\Box$
\section{The space $H^1(\cK(2),\cS \Psi \cD \cO (S^{1|2}) )$}

\subsection{The spectral sequence for a filtered module over a
Lie (super)algebra }
The reader should refer to \cite{P}, for the details of the
homological algebra used to construct spectral sequences. We will
merely quote the results for a filtered module $M$ with decreasing
filtration $\{M_n\}_{n\in\bbZ}$ over a Lie (super)algebra
$\mathfrak{g}$ so that $M_{n+1}\subset M_n,~\cup_{n\in \bbZ} M_n=M$
and $\mathfrak{g}M_n\subset M_n$ .

Consider the natural filtration induced on the space of cochains
by setting:
\begin{gather*}
F^n(C^*(\mathfrak{g},~M))=C^*(\mathfrak{g},~M_n),
\end{gather*}
then we have:
\begin{gather*}
  dF^n(C^*(\mathfrak{g},~M))\subset F^n(C^*(\mathfrak{g},~M))~~
  (\hbox{i.e.,
  the filtration is preserved by}~ d);\\
  F^{n+1}(C^*(\mathfrak{g},~M))\subset F^n(C^*(\mathfrak{g},~M))~
  \hbox{(i.e. the filtration is decreasing).}
\end{gather*}
Then there is a spectral sequence $(E_r^{*,*},d_r)$ for $r\in
\bbN$ with $d_r$ of degree $(r, 1-r)$ and
\begin{equation*}
E_0^{p,q}=F^p(C^{p+q}(\mathfrak{g},~M))/F^{p+1}(C^{p+q}(\mathfrak{g},~M))~~
\hbox{ and }~~ E_1^{p,q}=H^{p+q}(\mathfrak{g},~\Gr^p(M)).
\end{equation*}
To simplify the notations, we have to replace
$F^n(C^*(\mathfrak{g},~M))$ by $F^nC^*$. We define
\begin{gather*}
Z^{p,q}_r = F^pC^{p+q}\bigcap d^{-1}(F^{p+r}C^{p+q+1}),\\
B^{p,q}_r = F^pC^{p+q}\bigcap d(F^{p-r}C^{p+q-1}),\\ E^{p,q}_r =
Z^{p,q}_r/(Z^{p+1,q-1}_{r-1}+B^{p,q}_{r-1}).
\end{gather*}
The differential $d$ maps $Z^{p,q}_r$ into $Z^{p+r,q-r+1}_{r},$
and hence includes a homomorphism
\begin{equation*}
d_{r}:E^{p,q}_r\longrightarrow E^{p+r,q-r+1}_r
\end{equation*}
 The spectral sequence converges to $H^*(C,d),$
that is
\begin{equation*}
E_{\infty}^{p,q}\simeq F^{p}H^{p+q}(C,d)/F^{p+1}H^{p+q}(C,d),
\end{equation*}
where $F^{p}H^*(C,d)$ is the image of the map
$H^*(F^pC,d)\rightarrow H^*(C,d)$ induced by the inclusion
$F^pC\rightarrow C$.
\subsection{Computing $H^1(\cK(2),\cS \Psi \cD\cO(S^{1|2}))$}
Now we can check the behavior of the cocycles $\Upsilon_1,\ldots,
\Upsilon_{10}$ under the successive differentials of the spectral
sequence. Cocycles $\Upsilon_1,~\Upsilon_2$ and $\Upsilon_3$
belong to $E_1^{-1,2}$, cocycles $\Upsilon_4,\ldots, \Upsilon_9$
belong to $E_1^{0,1}$ and cocycle $\Upsilon_{10}$ belongs to
$E_1^{1,0}$. Consider a cocycle in $\cS\cP(2)$, but compute its
differential as if it were with values in $\cS \Psi \cD\cO
(S^{1|2})$ and keep the symbolic part of the result. This gives a
new cocycle of degree equal to the degree of the previous one plus
one, and its class will represent its image under $d_1$. The
higher order differentials $d_r$ can be calculated by iteration of
this procedure, the space $E^{p+r,q-r+1}_r$ contains the subspace
coming from $H^{p+q+1}(\cK(2);~\Gr^{p+1}(\cS \Psi
\cD\cO(S^{1|2})))$.

It is now easy to see that the cocycles
$\Upsilon_1,\ldots,\Upsilon_6$ will survive in the same form.
Computing supplementary higher order terms for the other cocycles,
we obtain
\begin{thm}
\label{th2} The space $H^1(\cK(2),\cS \Psi \cD\cO(S^{1|2}))$ is
purely even. It is spanned by the classes of the following
nontrivial $1$-cocycles
\renewcommand{\arraystretch}{1.4}
$$
\begin{array}{lll}
\Theta_1(v_F) =&\eta_1\eta_2(F)\xi^{-1}\zeta_1\zeta_2,\\
 \Theta_2(v_F) =&F'\xi^{-1}\zeta_1\zeta_2,\\
 \Theta_3(v_F) =&\Big(\displaystyle\frac{1}{4}(F+(-1)^{p(F)+1}F)+
 \eta_2\eta_1(F\theta_1\theta_2)\Big)\xi^{-1}\zeta_1\zeta_2,\\
 \Theta_4(v_F) =&\displaystyle
\frac{1}{4}(F+(-1)^{p(F)+1}F)+\eta_2\eta_1(F\theta_1\theta_2),\\
\Theta_5(v_F) =&\displaystyle F',\\
\Theta_6(v_F)=&\eta_1\eta_2(F),\\
\Theta_7(v_F)=&\displaystyle\sum_{n=0}^{\infty}
\frac{(-1)^{p(F)+n}}{n+1}\Big(\eta_1(F^{(n+1)})\zeta_1 +
\eta_2(F^{(n+1)})\zeta_2\Big)\xi^{-n-1}+\\ &\sum_{n=0}^{\infty}
\frac{2(-1)^{n}}{n+2} F^{(n+2)}\xi^{-n-1},\\
\Theta_{8}(v_F)=&\displaystyle \sum_{n=0}^{\infty}
(-1)^{p(F)+n}\Big(\eta_2(F^{(n+1)})\zeta_1 -
\eta_1(F^{(n+1)})\zeta_2\Big)\xi^{-n-1} + \\
&\sum_{n=0}^{\infty}(-1)^{n}F^{(n+2)}\xi^{-n-2}\zeta_1\zeta_2 +\\
&\sum_{n=1}^{\infty}(-1)^{n}\eta_1\eta_2(F^{(n)})\xi^{-n},\\
\Theta_{9}(v_F)=&\displaystyle \sum_{n=0}^{\infty}
(-1)^{n}\eta_1\eta_2(F^{(n+1)})\xi^{-n-2}\zeta_1\zeta_2 +\\
 &\sum_{n=1}^{\infty}
(-1)^{p(F)+n}\frac{n}{n+1}\Big(\eta_1(F^{(n+1)}) \zeta_1
+\eta_2(F^{(n+1)})\zeta_2\Big)\xi^{-n-1}+\\
 &\sum_{n=1}^{\infty}(-1)^{n}\frac{n}{n+2}F^{(n+2)}\xi^{-n-1},\\
\Theta_{10}(v_F)=&\displaystyle \sum_{n=1}^{\infty}
(-1)^{n+1}\frac{2n}{n+2}F^{(n+2)}\xi^{-n-2}\zeta_1\zeta_2+\\
  &\sum_{n=1}^{\infty} (-1)^{p(F)+n}\frac{2n}{n+1}
  \eta_1(F^{(n+1)})\xi^{-n-1}\zeta_2 +\\
  &\sum_{n=1}^{\infty}(-1)^{p(F)+n+1}\frac{2n}{n+1}
  \eta_2(F^{(n+1)})\xi^{-n-1}\zeta_1 +\\
 &\sum_{n=1}^{\infty}2(-1)^{n+1}\eta_1\eta_2(F^{(n)})\xi^{-n}.
\end{array}
$$
\end{thm}
\bigskip

\noindent {\bf Acknowledgements} It is a pleasure to thank
Valentin Ovsienko who introduced us to the question of cohomology
computations in Lie superalgebras of vector fields. We also thank
Dimitry Leites and Claude Roger for helpful discussions.


\label{lastpage}


\begin{thebibliography}{99}
\bibitem{Ra}
A.O.Radul, Non-trivial central extensions of Lie algebras of
differential operators in two higher dimensions, Phys. Letters B
265 (1991) 86--91.

\bibitem{AB}
B.Agrebaoui, N.Ben Fraj, {\it On the cohomology of the Lie
Superalgebra of contact vector fields on $S^{1|1}$}, Belletin de
la Soci\'et\'e Royale des Sciences de Li\`ege, vol. 72, 6, 2004,
365--375.

\bibitem{Fu}
D. B. Fuks, {\it Cohomology of infinite-dimensional Lie algebras},
Plenum Publ. New York, 1986.

\bibitem{BO}
N. Ben Fraj, S. Omri, {\it Deforming the Lie Superalgebra of
Contact Vector Fields on $S^{1|1}$ inside the Lie Superalgebra of
Superpseudodifferential operators on $S^{1|1}$,} J. Nonlinear
Mathematical Physics (to appear).

\bibitem{RO1}
V. Ovsienko \& C. Roger, {\it Deforming the Lie algebra of vector
fields on~$S^1$ inside the Lie algebra of pseudodifferential
operators on $S^1$}, AMS Transl. Ser.~2, (Adv. Math. Sci.)
vol.~194 (1999) 211--227.

\bibitem{P}
E. Poletaeva, The analogs of Riemann and Penrose tensors on
supermanifolds, arXiv: math.RT/0510165.
\end{thebibliography}
\end{document}